\documentclass[%
 reprint,
runinaddress,
 amsmath,amssymb,
aps,
prb,
showkeys
]{revtex4-1}

\usepackage{graphicx}
\usepackage{hyperref}
\usepackage{dcolumn}
\usepackage{bm}
\usepackage[utf8]{inputenc} 
\usepackage{float}
\usepackage{xcolor}

\pdfoutput=1

\begin{document}

\bibliographystyle{unsrtnat} 

\title{Quantum spill out in few-nanometer metal gaps: Effect on gap plasmons and reflectance from ultrasharp groove arrays}%

\author{Enok J. H. Skjølstrup}
\email{ejs@mp.aau.dk}
\author{Thomas Søndergaard}
\author{Thomas G. Pedersen}
\affiliation{%
 Department of Materials and Production, Aalborg University, Skjernvej 4A, DK-9220 Aalborg East, Denmark. 
}%

\date{\today}

\begin{abstract}
Plasmons in ultranarrow metal gaps are highly sensitive to the electron density profile at the metal surfaces. Using a fully quantum mechanical approach, we study the effects of electron spill-out on gap plasmons and reflectance from ultrasharp metal grooves. We demonstrate that the mode index of ultranarrow gap plasmons converges to the bulk refractive index in the limit of vanishing gap and, thereby, rectify the unphysical divergence found in classical models. Surprisingly, spill-out also significantly increases the plasmonic absorption for few-nanometer gaps and lowers the reflectance from arrays of ultrasharp metal grooves. These findings are explained in terms of enhanced gap plasmon absorption taking place inside the gap 1-2 Å from the walls and delocalization near the groove bottom. Reflectance calculations taking spill-out into account are shown to be in much better agreement with measurements compared with classical models.
\end{abstract}

\maketitle

\section{Introduction} \label{sec:introduction}
In the past decade, plasmonic structures have attracted attention, in part due to their efficiency in absorbing incident light \cite{lalanne1,lalanne2} and their ability to squeeze light below the diffraction limit \cite{control1,control2,control3,control4}. Furthermore, plasmonic structures can be applied in e.g. solar cells, lasers, and biosensors \cite{application1,application2,application3}. Metal surfaces support surface plasmon polaritons (SPPs) that are electromagnetic waves bound to and propagating along the surface, while deep-subwavelength gaps between metal surfaces may support gap plasmons, i.e. waves confined to and propagating along the gap. 
Such gap plasmons localized in gaps of nanometer size between spherical and triangular nanoparticles have been studied in Refs. \citenum{control1,narrow_gap2}. Furthermore, the propagation of gap plasmons in wider gaps between two parallel metal surfaces has been studied in Refs. \citenum{narrow_gap1,NJP,gap1,gap2}, and when propagating in rectangular or tapered grooves in Refs. \citenum{gap3,gap4,extinction}. 
In all these papers, quantum spill-out is neglected, such that the dielectric function takes one value in the gap region and another value in the metal, thus changing abruptly at the interfaces. 

In this paper, we focus on gaps of a few nm in metals, which can be found in ultrasharp groove arrays \cite{NJP,GSP,optics_multiple,black_gold,narrow_gap3}. In such grooves, nearly parallel metal surfaces are separated by an ultranarrow gap near the bottom. These structures are broadband absorbers of light\cite{optics_multiple,black_gold}, and most of the absorption takes place in the bottom part of the grooves \cite{optics_single}. In addition, most of the physics can be explained in terms of gap plasmons propagating back and forth in the grooves. So far, the modeling of ultrasharp grooves has not taken quantum spill-out effects into account, and in the extreme limit of vanishing gap width, the resulting gap plasmon mode index diverges \cite{NJP,gap1,gap2,gap3,gap4}, which is clearly unphysical. Importantly, only minor oscillations are observed in the measured reflectance spectra from Ref. \citenum{black_gold}, which does not match present theories neglecting spill-out. 

We show in this paper that taking quantum spill-out into account leads to a drastically improved agreement with the measured reflectance from arrays of ultrasharp grooves in gold films. Furthermore, the mode index of gap plasmons converges to the refractive index of bulk gold for vanishing gaps, thus restoring physically correct behaviour. The range of electron spill-out is only about 0.3 nm, implying that when the gap width is below 0.6 nm the electron distributions from the two gold surfaces overlap, and electrons can tunnel across the gap, while the surfaces do not couple electronically for wider gaps. Surprisingly, we demonstrate in the following that the effect of spill-out also significantly increases the absorption when the gap width is a few nm, thus far outside the tunnel regime.

\section{Quantum dielectric function} \label{sec:density}
In a quantum mechanical description of metal surfaces, the electron density has an exponential tail stretching into the vacuum region due to tunneling through the surface barrier. A highly efficient model of such spill-out effects is provided by Density-Functional Theory (DFT) in the jellium approximation treating the positive ions as a constant charge density inside the metal \cite{jellium1,jellium2}. The distribution of free ($s,p$ band) electrons in this positive background produces an inhomogeneous electron density with a characteristic spill-out into the vacuum region. 

The density of free electrons in the vicinity of a gap between metal surfaces is found by self-consistently solving the Kohn-Sham equations \cite{jellium1} within the jellium model. The exchange and correlation potentials that appear in these equations are calculated applying the Local Density Approximation (LDA)\cite{kohanoff}, using the Perdew-Zunger parametrization in the correlation term \cite{perdew}. The applied Wigner-Seitz radius for gold is $r_s=3.01$ Bohr \cite{jellium2}. The electron density is calculated for a structure consisting of two parallel gold slabs of width $d$ separated by the gap width $w$. The width of the slabs must be thick enough such that artificial finite-size effects are negligible, and we find that $d=2.6$ nm is sufficient. The density is calculated using a standing-wave basis of the form $\sin(n\pi(x/L+1/2))$, where $n=1, 2, \cdots, 200$ and $L=8$ nm, and it has been checked that the density has converged with respect to the number of basis functions. A variation in Fermi energy between two iterations below $10^{-7}$ Ha is used as convergence criterion in the optimization process, and an Anderson mixing scheme \cite{anderson} with mixing parameter 0.005 is applied.

Such a DFT model has in other cases previously been applied to calculate the optical cross sections of metal nanowires \cite{nanowires1,nanowires2,nanowires3,nanowires4}, metal clusters and spheres \cite{clusters1,clusters2,nanowires1,response1}, and the plasmon resonance of metal dimers \cite{pol, response2} and of semiconductor nanocrystals \cite{response3}. In addition, it was shown in Ref. \citenum{abajo} that electron spill-out has a significant impact on the local density of states for gap plasmons propagating between two gold surfaces. 
   \begin{figure} [ht]
   \begin{center}
   \begin{tabular}{c} 
   \includegraphics[width=8cm]{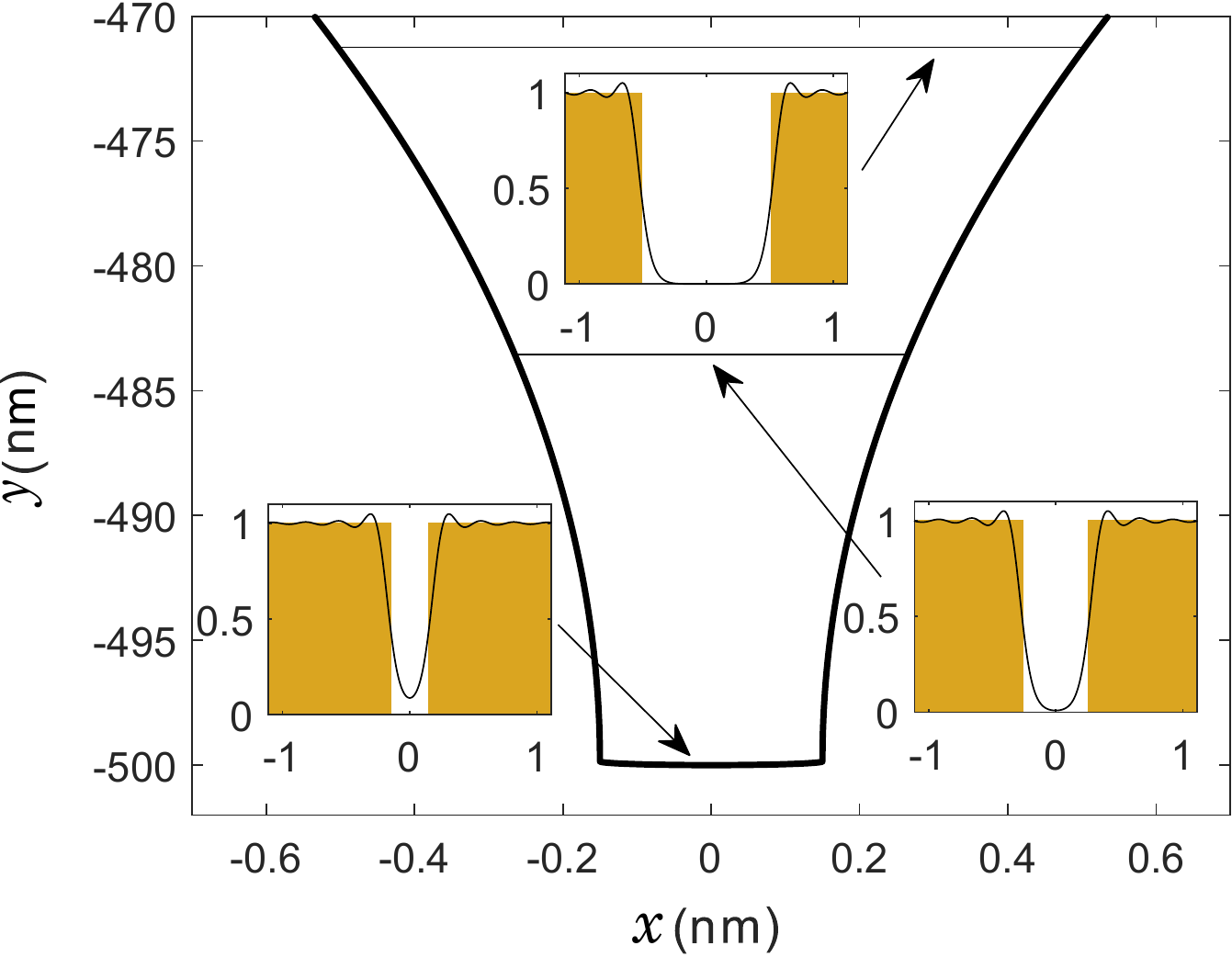}
   \end{tabular}
   \end{center}
   \caption[example] 
   { \label{fig:density} 
Schematic of the bottom 30 nm of an ultrasharp groove with a bottom width of 0.3 nm. The three insets show the electron density $n/n_0$ at the bottom of the groove and for gap widths of 0.5 nm and 1 nm. The colored areas in the insets show the position of the gold surfaces. }
   \end{figure} 
   
A continuous range of widths between metal surfaces can be found in ultrasharp grooves \cite{NJP,black_gold,optics_multiple}. Fig. \ref{fig:density} shows a schematic of the bottom 30 nm of a groove with bottom width of 0.3 nm in accordance with the geometry considered in Refs. \citenum{black_gold,optics_multiple}. The three insets show the calculated electron density $n(x,w)$ in units of the bulk gold electron density $n_0$ at three different cross-sections of the ultrasharp groove corresponding to gap widths $w$ = 0.3, 0.5, and 1 nm, respectively. The colored areas mark the position of the gold surfaces. Due to spill-out, the electron density extends a small distance into the gap and, especially for the smallest gap width of 0.3 nm, the density only decreases to roughly 9\% of the bulk gold density in the center of the gap, while it decreases to about 1\% when the gap width is 0.5 nm. Hence, there is no
true vacuum region between the gold surfaces for these gap widths. 
For the larger gap of 1 nm, the electron density drops practically to zero 0.3 nm from the gold surfaces, such that the two gold surfaces do not couple electronically. The range of spill-out is therefore 0.3 nm, and electrons can tunnel from one gold surface to the other only when the gap width is below 0.6 nm. 
In addition, the electron density inside the metal is also affected near the surface and shows Friedel oscillations, in agreement with previous studies of the electron density across a single boundary between metal and air\cite{jellium1,jellium2,yan}. 

It is noticed that all the calculated densities only include spill-out from parallel gold slabs, and these densities are merged together in a region, where the curvature of the groove walls is small in order to form the density across the two-dimensional ultrasharp groove. Electron spill-out also occurs from the bottom of the groove and affects the density on a length scale comparable to the spill-out range. However, in this paper we ignore spill-out from the bottom, an assumption that will be explained in Sec. \ref{sec:reflectance}.

Next, the electron density is applied to compute the dielectric function $\varepsilon$ across the structure as described by the Drude model modified to include the inhomogeneous density. In addition to the free electrons, bound electrons are found in lower lying $d$ bands, and they contribute to the interband part of the dielectric function \cite{nanooptik}. 
In contrast to the free electrons, we assume no spill-out of bound electrons into the gap region, and they are therefore entirely located in the bulk region. In Refs. \citenum{quantal1,quantal2,quantal3}, a thin surface layer with ineffective screening has been applied to account for the interband part of the dielectric function, such that this contribution is a step function changing abruptly a few Å from the surface at the metal side of the interface. Applying this model, plasmon resonances in nanometer-size clusters of gold, silver, and copper are calculated and are in good agreement with measurements. In this paper, however, similar to Ref. \citenum{abajo}, we ignore this thin surface layer and assume thereby that the interband part of the dielectric is a step function changing abruptly at the interfaces between air and gold, thus at the same position as the jellium edge. It is shown in Sec. \ref{sec:reflectance} that this simple description of the bound electrons leads to good agreement with measurements of the reflectance of an ultrasharp groove array. 

In the bulk, the electron density $n_0$ implies a bulk plasma frequency of $\omega_{p,\textrm{bulk}}=\sqrt{n_0e^2/(m_e \varepsilon_0)}$ and an accompanying Drude response $\varepsilon_{p,\textrm{bulk}}(\omega)=1-\omega_{p,\textrm{bulk}}^2/(\omega^2+i\omega\Gamma)$. We include the interband contribution by requiring the bulk response to equal the measured dielectric function $\varepsilon_{\textrm{gold}}(\omega)$ from Ref. \citenum{christy} so that the final dielectric function is given by
\begin{align}
\varepsilon(\omega,x,w)&=1-\frac{\omega_p^2(x,w)}{\omega^2+i\Gamma \omega} \nonumber  \\
&+\left[\varepsilon_{\textrm{gold}}(\omega)-\varepsilon_{p,\textrm{bulk}}(\omega)\right]\theta(\vert x \vert-w/2). \label{eq:eps}
\end{align}
Here, the first term describes the Drude response of free electrons with plasma frequency $\omega_p(x,w)=\sqrt{n(x,w)e^2/(m_e \varepsilon_0)}$ determined by the electron density $n(x,w)$ calculated using DFT. In addition, the damping term in gold is $\hbar \Gamma=65.8$ meV \cite{nanooptik} and the step function models the abrupt behavior assumed for the bound electron interband term.

\begin{figure}[!h]
\begin{center}
\includegraphics[width=8cm]{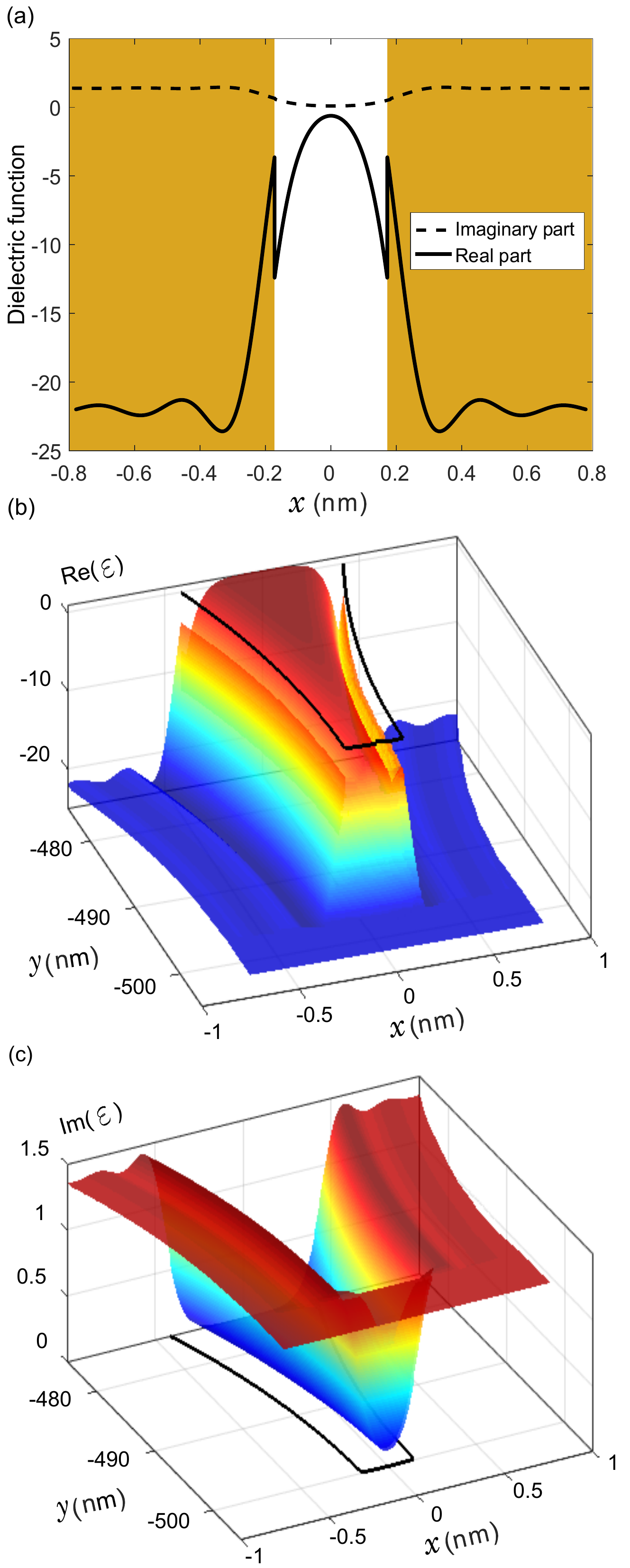}
\end{center}
\caption{(a) Real and imaginary part of $\varepsilon(x,w)$ for $w=0.35$ nm, where the colored areas show the position of the gold surfaces. (b) and (c) 3D plots of real and imaginary part of $\varepsilon(x,y)$ in the bottom 25 nm of an ultrasharp groove. The wavelength is 775 nm. }
\label{fig:Eps}
\end{figure}
An example of the dielectric function is seen in Fig. \ref{fig:Eps}(a) for $w=0.35$ nm at a wavelength of 775 nm, where the colored areas mark the position of the gold surfaces. The real part is clearly seen to jump according to the step function in Eq. (\ref{eq:eps}) and the dielectric function equals the bulk value in the gold regions that are sufficiently far from the air-gold interfaces. The imaginary part in Fig. \ref{fig:Eps}(a) also jumps, but it is difficult to see in the figure. The Friedel oscillations in the electron density (Fig. \ref{fig:density}) result in corresponding oscillations in the dielectric function (Fig. \ref{fig:Eps}(a)).

Fig. \ref{fig:Eps}(b) shows a 3D plot of the real part of the dielectric function $\varepsilon(x,y)$ in the bottom 25 nm of the ultrasharp groove at wavelength 775 nm. The black curve in the $xy$-plane shows the structure of the groove. A 3D plot of the imaginary part of  $\varepsilon(x,y)$ is seen in Fig. \ref{fig:Eps}(c). The imaginary part is small but non-zero in the entire shown region implying that absorption of light takes place even in the middle of the gap. 

As spill-out from the bottom of the groove is neglected there is pure gold at positions below $y=-500$ nm as illustrated by the blue color in Fig. \ref{fig:Eps}(b) and the red color in Fig. \ref{fig:Eps}(c). As $y$ increases and the groove width increases one should think that the effect of electron spill-out would become negligible as it only occurs very close to the groove walls as observed in Fig. \ref{fig:density}. 
However, most surprisingly, when it comes to the mode index and the absorption density, spill-out also has a great influence for gap widths of a few nm, even though they far exceed the tunnel regime.

\section{Mode index and absorption density of a propagating gap plasmon} \label{sec:modeindex}
A gap between two metal surfaces supports gap plasmons propagating in the $y$-direction. Gap plasmons are $p$-polarized electromagnetic waves, implying that the corresponding magnetic field  $\vec{H}(\vec{r})=\hat{z}H(x,y)$ for a constant $w$ is given by \cite{GSP}
\begin{equation}
H(x,y)=e^{ik_0\beta y}H(x), \label{eq:H}
\end{equation}
where $H(x)$ is the transverse field distribution, $\beta$ is the complex mode index, and $k_0=2\pi/\lambda$ is the free space wave number. Both the transverse field distribution and the mode index depend strongly on the gap width $w$.

The mode index is calculated for a fixed width $w$ by applying a transfer-matrix method \cite{optik}. This is done by dividing the $x$-axis into $N$ sufficiently thin layers, each modelled as having a constant dielectric function. For the mode index to converge it is found that thicknesses of $2.7 \cdot 10^{-4}$ nm are sufficient.
A structure matrix $\mathcal{S}$ is constructed, which relates the magnetic field to the left of the structure to that at the right of the structure
\begin{align}
\mathcal{S}=\mathcal{T}_{g1}\prod_{i=1}^{N-1}\left(\mathcal{T}_i\mathcal{T}_{i,i+1}\right)\mathcal{T}_N\mathcal{T}_{N,g}:=\begin{pmatrix} 
\mathcal{S}_{11} & \mathcal{S}_{12} \\ \mathcal{S}_{21} & \mathcal{S}_{22}
\end{pmatrix}.  \label{eq:S}
\end{align}
Here, a matrix with a single index denotes propagation in
that particular layer, a matrix with two indices denotes
an interface matrix, and the subscript $g$ denotes bulk gold. Expressions for the propagation and interface matrices can be found in Ref. \citenum{optik}.
The field to the left of the structure is then given by
\begin{align}
\begin{pmatrix} 
0 \\ H_L^- 
\end{pmatrix}=
\mathcal{S}
\begin{pmatrix} 
H_R^+ \\ 0 
\end{pmatrix}. \label{eq:S1}
\end{align}
Here, $R$ and $L$ denote right and left of the structure, respectively, and $+$ and $-$ denote the direction, in which light propagates. Left and right of the structure there is only light propagating in the negative and positive direction, respectively. From Eq. (\ref{eq:S1}) it is found that the matrix element $\mathcal{S}_{11}$ must be zero. The matrices in Eq. (\ref{eq:S}) depend on the mode index $\beta$, implying that $\mathcal{S}_{11}$ is a function of $\beta$, and to find the roots $\mathcal{S}_{11}$ is evaluated for a range of complex $\beta$ values. By observing sign changes in the real and imaginary parts of $\mathcal{S}_{11}$ a region of $\beta$ values is identified where $\mathcal{S}_{11}$ is close to zero. The Newton-Raphson method is then applied to obtain the exact root.
Four periods of Friedel oscillations are found sufficient in the modelling of the dielectric function near the gold surface in order for the mode index to converge, c.f. Fig. \ref{fig:Eps}(a), and the bulk value is applied beyond this range. 

\begin{figure}[!h]
\begin{center}
\includegraphics[width=8cm]{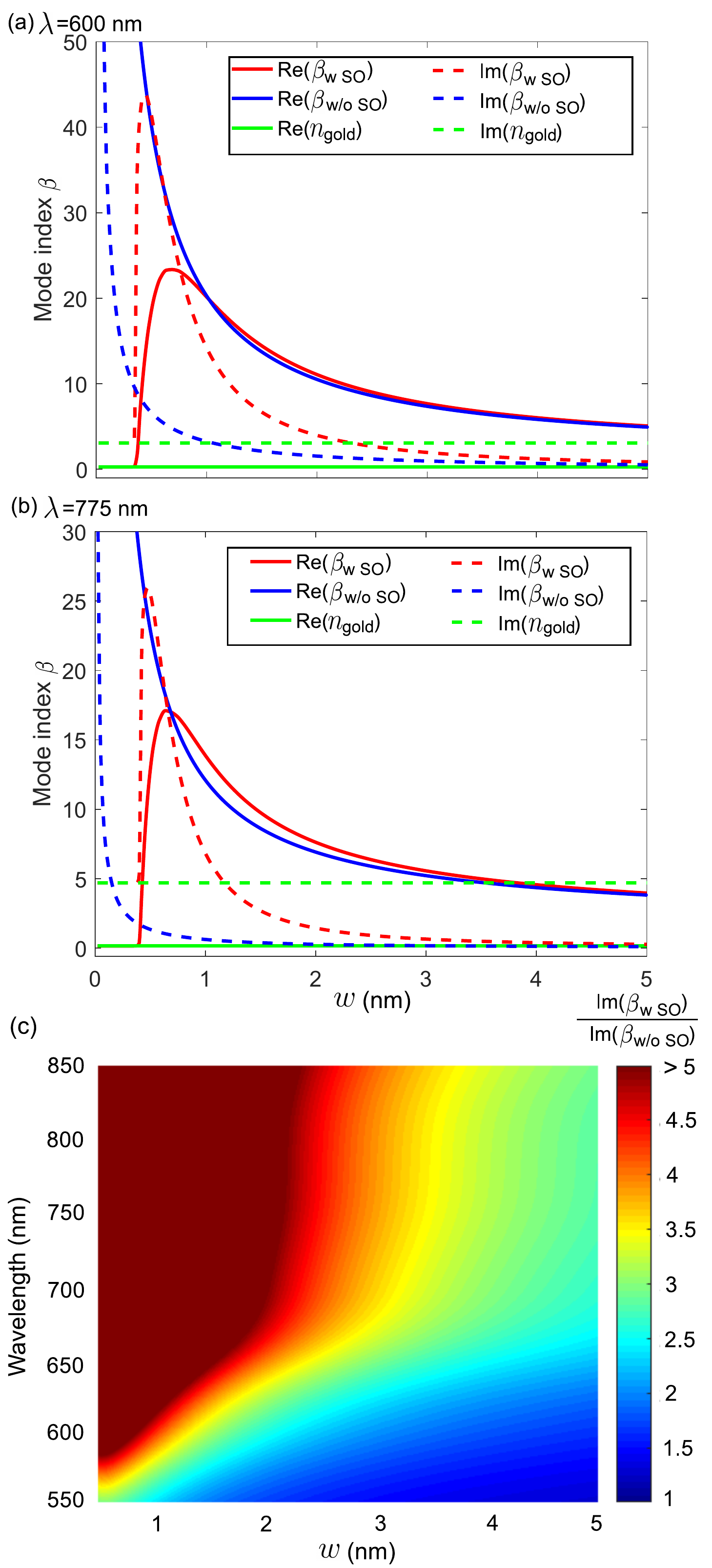}
\end{center}
\caption{(a,b) Real part (solid red line) and imaginary part (dashed red line) of the mode index of a gap plasmon as a function of the gap width $w$ when taking electron spill-out into account. The horizontal green lines show the refractive index of bulk gold. The blue lines show the corresponding mode index when neglecting spill-out. The wavelength is 600 nm in (a) and 775 nm in (b). (c) Ratio between imaginary parts of the mode index with and without spill-out (SO).}
\label{fig:modeindex}
\end{figure}

Applying this method the mode index of a gap plasmon has been calculated using the dielectric function from the previous section. The mode index as a function of $w$ is shown in Fig. \ref{fig:modeindex}(a) for $\lambda=600$ nm and in Fig. \ref{fig:modeindex}(b) for $\lambda=775$ nm. 

In Ref. \citenum{abajo} gap plasmons propagating between two gold surfaces are also studied when including spill-out. In that paper in particular the local density of states is calculated, from which it is possible to obtain a dispersion relation. It shows the photon energy as a function of parallel wave number (analogue to $k_0\beta$ in this paper) for some gap width. But there only the real part of the wave number is considered, and in addition it is not shown how this real part explicitly depends on the gap width $w$. But in this paper, we calculate both components of the complex mode index exactly, and show in Fig. \ref{fig:modeindex}(a,b) how they are explicitly dependent on $w$.

The red solid and dashed lines are the real and imaginary parts of the mode index obtained when including electron spill-out. The horizontal green lines represent the real and imaginary part of the refractive index $n_{\textrm{gold}}$ from Ref. \citenum{christy}, and the mode index is seen to converge to this value for sufficiently small $w$. It is difficult to see in the figure, but the real part of $n_{\textrm{gold}}$ shown by the solid green line is positive but very small.
The blue lines show the mode index obtained when neglecting electron spill-out similar to previous studies \cite{NJP,gap1,gap2}. Here, the mode index diverges for $w \to 0$, which cannot be correct from a physical point of view. The mode index must converge to the refractive index of bulk gold for small gaps, since in this case the structure is simply bulk gold. For distances $w$ below 0.35 nm, the mode index of the gap plasmon is almost the same as the refractive index of pure gold. Hence, even though the dielectric function for $w$ = 0.35 nm is quite different from that of pure gold (see Fig. \ref{fig:Eps}(a)), the system nevertheless behaves almost as pure gold when it comes to the mode index of a propagating gap plasmon. For large $w$, the mode index both with and without spill-out converges to $\sqrt{\epsilon_{\textrm{gold}}/(\epsilon_{\textrm{gold}}+1)}$ \cite{nanooptik}, as in this case the wave behaves as an SPP bound to a single interface (not shown).

On the other hand, when the gap width is a few nm, thus far outside the tunnel regime, the real part of the mode index is almost the same with and without spill-out, as seen by comparing the red and blue solid lines in Fig. \ref{fig:modeindex}(a,b). However, there is a large difference between the imaginary parts of the mode index, as seen by comparing the corresponding dashed lines. This is further illustrated in Fig. \ref{fig:modeindex}(c) showing the ratio of the imaginary parts of the mode index with and without spill-out (SO). For gaps of a few nm, the imaginary part of the mode index is seen to be much higher when including spill-out, especially for long wavelengths. In the figure, the color is dark red for every value above 5, but the maximal value is more than 20 which is found for a gap width of 0.5 nm at a wavelength of 850 nm. Hence for few-nanometer gaps the effect of spill-out on the imaginary part of the mode index is significant. 

\begin{figure}[!h]
\begin{center}
\includegraphics[width=8cm]{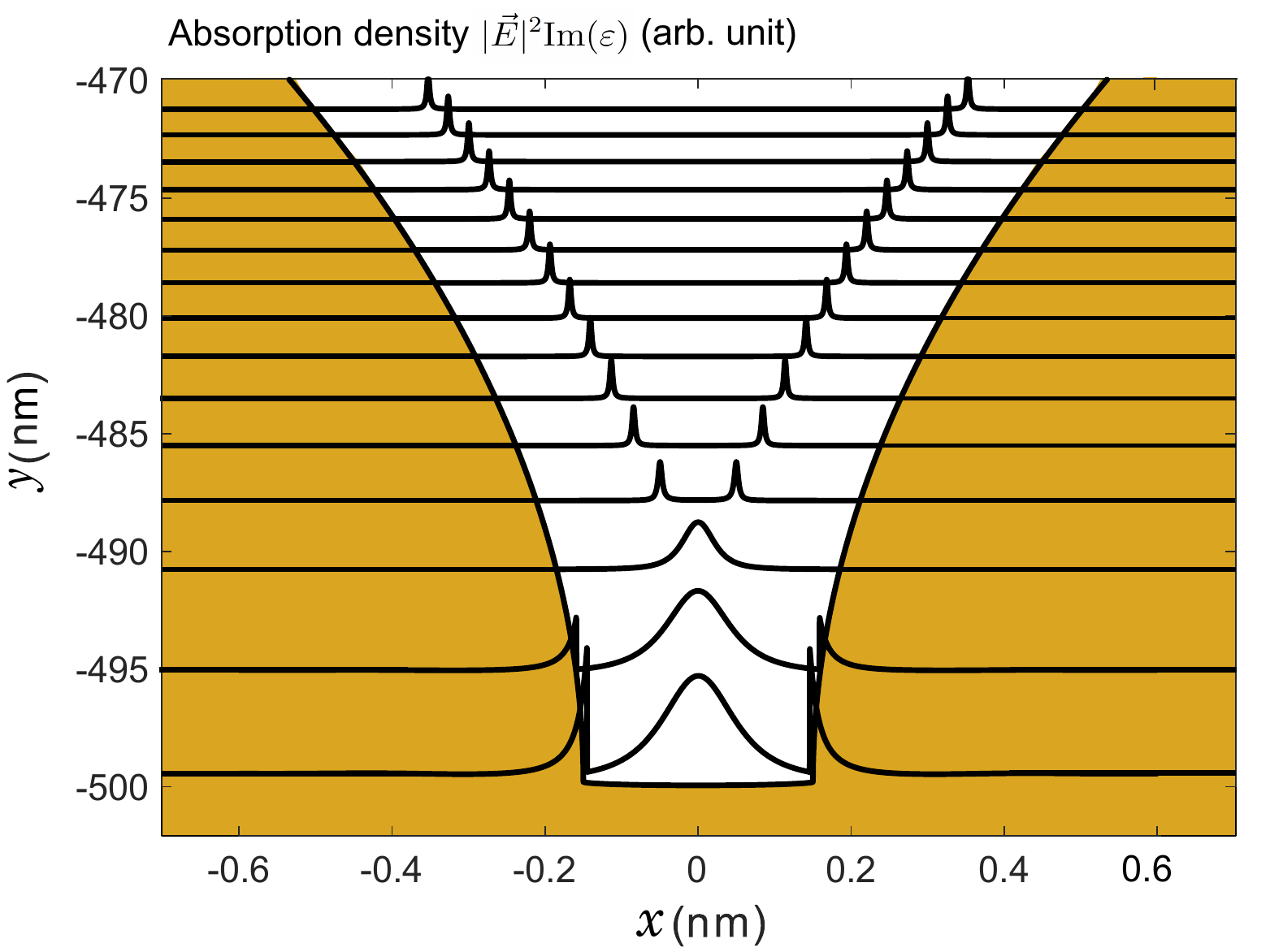}
\end{center}
\caption{(Absorption density in the bottom 30 nm of an ultrasharp groove. In the bottom 10 nm absorption takes place across the entire gap, but as $y$ increases and the groove gets broader, absorption mostly takes place 0.15 nm from the interfaces. The wavelength is 775 nm.}
\label{fig:abs_density}
\end{figure}

To elucidate the physics behind the increased imaginary part, the absorption density 
\begin{align}
A(x,y)=\vert \vec{E}(x,y)\vert^2 \textrm{Im}(\varepsilon(x,y))
\end{align}
is calculated as shown in Fig. \ref{fig:abs_density} in the bottom 30 nm of an ultrasharp groove at a wavelength of 775 nm, where the colored areas show the position of the gold. 
Here, the electric field has been calculated from the magnetic field in Eq. (\ref{eq:H}) as \cite{nanooptik}
\begin{align}
\vec{E}=\frac{i}{\omega\varepsilon_0\varepsilon}\vec{\nabla}\times \hat{z}H. \label{eq:E}
\end{align}
In the bottom of the groove, where the gap width is 0.3 nm, the absorption density clearly jumps across the boundaries in order for the normal part of the displacement field to be continuous across the interfaces \cite{nanooptik}. It is noticed that the dielectric function on the gap side of the interface has a numerically higher value than on the gold side of the interface, which is due to the abrupt jump in the bound electron term (see Fig. \ref{fig:Eps}(a)). In order for the displacement field to be continuous, this implies that the electric field magnitude is correspondingly lower on the gap side than on the gold side, which explains why the absorption density in the bottom of Fig. \ref{fig:abs_density} increases across the interface. If spill-out is neglected, the dielectric function has a numerically higher value on the gold side of the interface, which implies that in this case the electric field magnitude drops across the interface.  
As $y$ increases and the groove gets broader, the absorption density mostly consists of two peaks located about 0.15 nm from the groove walls. At these positions, the real part of the dielectric function is zero (at the wavelength 775 nm), while the imaginary part is small but non-zero.

When neglecting spill-out the dielectric function has zero imaginary part in the gap and absorption can only take place in the metal. It is highly surprising that the effect of spill-out significantly increases the absorption density also for few-nm gaps that far exceed the tunnel regime. 
How this affects the reflectance from an ultrasharp groove array is studied in the next section.

\section{Reflectance from an ultrasharp groove array} \label{sec:reflectance}
Periodic arrays of ultrasharp grooves absorb light almost perfectly in a broad wavelength interval, which is utilized in plasmonic black gold, where the grooves turn a shiny gold surface into a broadband absorber \cite{black_gold,optics_multiple}. Rectangular or tapered (not ultrasharp) grooves may on the other hand absorb efficiently in a narrow band of wavelengths \cite{Miyazaki,Greffet,optics_rectangular,extinction}, which is advantageous in thermophotovoltaics \cite{bauer,sai}. Hence, the reflectance spectrum of an array of grooves strongly depends on the groove shape. Previously the reflectance of a groove array has only been calculated when neglecting spill-out, but in this section it is calculated when taking spill-out into account. This is done by applying the stack matrix method (SMM) of Ref.\citenum{GSP}, where the grooves are divided into layers with a refractive index corresponding to the gap plasmon mode index. In Ref. \citenum{GSP}, the SMM was applied to calculate reflectance from ultrasharp groove arrays when neglecting spill-out, and the results were practically identical to results obtained using a full Greens function surface integral equation method. 

As mentioned in the previous section, the mode index of the propagating gap plasmon is almost the same as in pure gold for gap widths below 0.35 nm (see Fig. \ref{fig:modeindex}(a,b)) even though the dielectric function is quite different from that of pure gold (see Fig. \ref{fig:Eps}(a)). By looking at the groove structure in Fig. \ref{fig:density}, it is found that the distance between the groove walls is below 0.35 nm at the bottom 8 nm of the groove, thus the gap plasmon behaves almost as pure gold in this region. As mentioned in Sec. \ref{sec:density} electron spill-out also occurs from the bottom and affects the density in a short range from the bottom. But, as the range of spill-out is much shorter than the 8 nm, it only influences the dielectric function at positions where the mode index already behaves almost as pure gold and where minor variations in dielectric function implies no change in mode index. This explains why we have reasons to ignore the spill-out from the bottom of the groove. 

\begin{figure}[!h]
\begin{center}
\includegraphics[width=8cm]{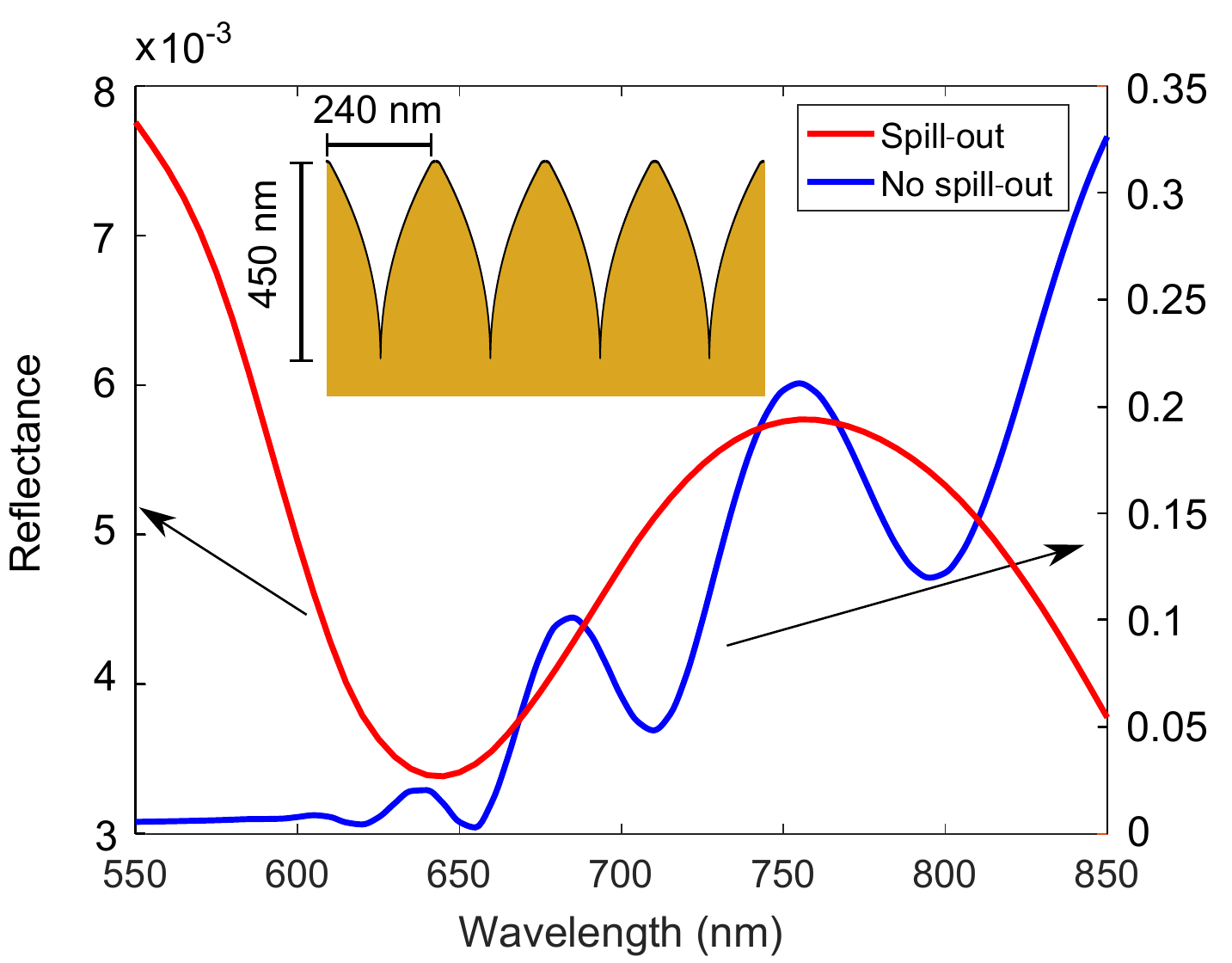}
\end{center}
\caption{Reflectance from an ultrasharp groove array in gold where the grooves have a top width of 240 nm, a bottom width of 0.3 nm, and a groove height of 450 nm. The grooves are illuminated by normally incident light. Spill-out is included in the calculated reflectance shown by the red line on the left $y$-axis, and neglected in the corresponding reflectance shown by the blue line on the right $y$-axis. The inset shows a schematic of the groove geometry.}
\label{fig:R}
\end{figure}

The array of ultrasharp grooves in gold is illustrated in the inset in Fig. \ref{fig:R}. 
Here the reflectance from an ultrasharp groove array illuminated by normally incident light is shown for the wavelength interval 550-850 nm, where the groove height is 450 nm, the top width is 240 nm, and the bottom width is 0.3 nm. Including spill-out leads to the reflectance shown by the red line on the left $y$-axis in Fig. \ref{fig:R}. The same groove dimensions were applied in Ref. \citenum{black_gold} where electron spill-out was neglected, which gives the reflectance shown by the blue line on the right $y$-axis in the same figure. It is clearly seen that the effect of spill-out significantly lowers the reflectance from an ultrasharp groove array in gold. This is expected since the imaginary part of the gap plasmon mode index is higher when including spill-out (see Fig. \ref{fig:modeindex}(c)). 

The mode index and, thus, the degree of absorption depends on how the field profile is distributed between the gap and the metal regions. It is therefore investigated how spill-out affects the field profile for a small gap.
The electric field of the gap plasmon in Eq. (\ref{eq:E}) has both an $x$- and a $y$-component, where the $x$-component jumps across the interface as observed in the bottom of Fig. \ref{fig:abs_density}. The corresponding magnetic field in Eq. (\ref{eq:H}) only has a $z$-component, which is continuous across the interface \cite{nanooptik}, and makes the magnetic field preferable for illustrating the penetration of the field into the metal. 

\begin{figure}[!h]
\begin{center}
\includegraphics[width=8cm]{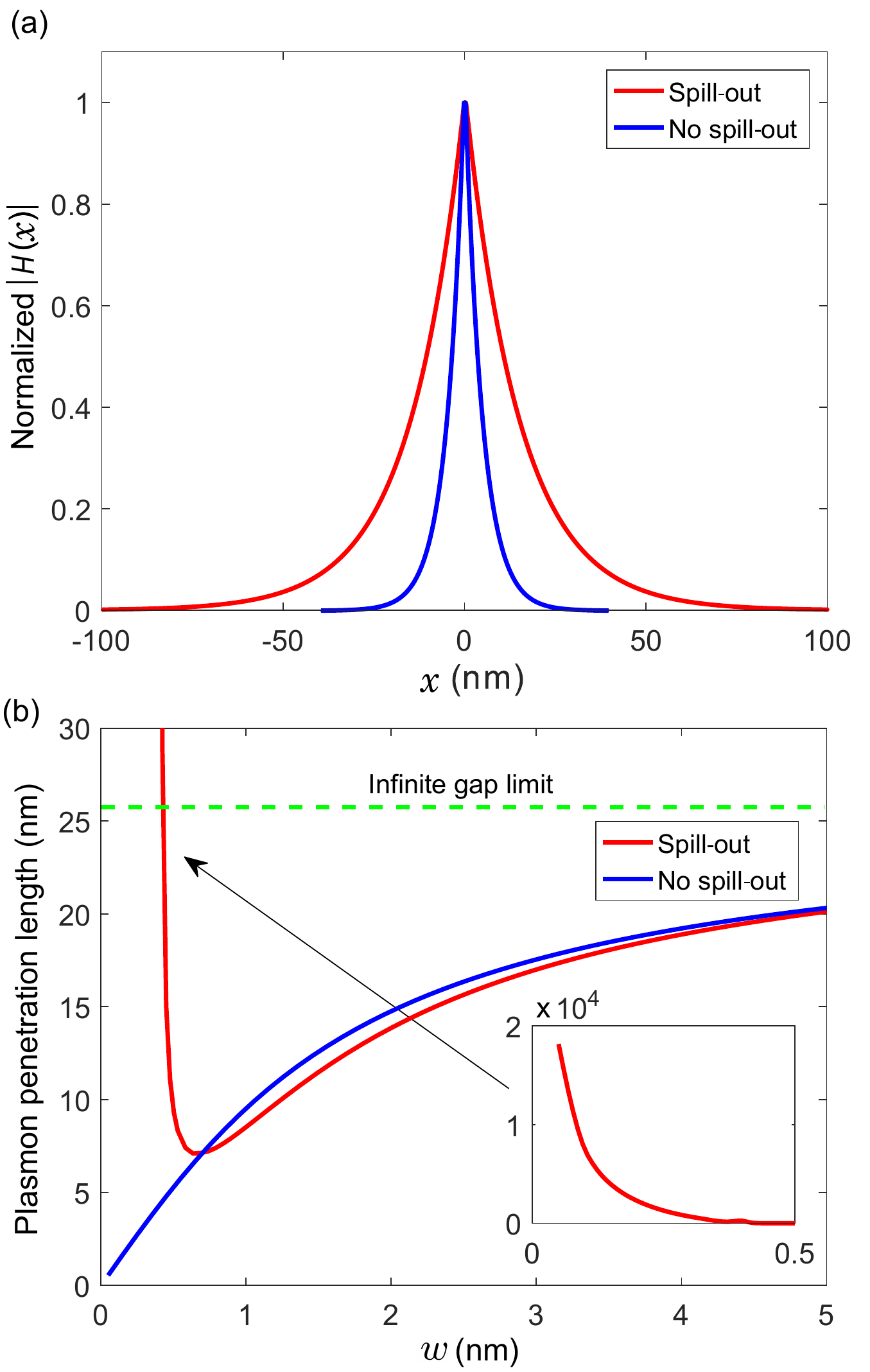}
\end{center}
\caption{(a) Normalized $\vert H(x) \vert$ for a gap plasmon propagating in a gap of width $w=0.45$ nm. Spill-out is included and neglected in the calculations of the fields shown by the red and blue line, respectively. (b) Corresponding plasmon penetration length as a function of $w$ with and without spill-out. Horizontal green line shows the penetration length of an SPP bound to a single gold surface, i.e in the limit of an infinite gap. The wavelength is 775 nm in both (a) and (b).}
\label{fig:H}
\end{figure}

The gap plasmon transverse field magnitude $\vert H(x) \vert$ is shown in Fig. \ref{fig:H}(a) for a gap width of $w=0.45$ nm both with and without spill-out at a wavelength of 775 nm. The field distributions have been normalized by their maximum value. 
The blue line shows the field when neglecting spill-out. In this case the mode index  $\beta$ is high (see blue lines in Fig. \ref{fig:modeindex}(b)), which implies that the imaginary part of $k_x=k_0\sqrt{n^2_{\textrm{gold}}-\beta^2}$ is also high. The field is therefore partly localized in the air gap region and the penetration length calculated as $1/\textrm{Im}(k_x)$ is only 4.8 nm.

When including spill-out for the same $w$ the mode index is closer to that of pure gold (see red lines in Fig. \ref{fig:modeindex}(b)), and $k_x$ is therefore smaller, which yields a more delocalized field as shown by the red line in Fig. \ref{fig:H}(a). Here the penetration length is 15 nm, and most of the field is therefore located in the pure gold regions. But as the absorption density for this $w$ consists of two peaks, as in the upper part of Fig. \ref{fig:abs_density}, absorption mostly takes place 0.15 nm from the interfaces and not in the pure gold regions, even though the magnetic field in Fig. \ref{fig:H}(a) is mostly located there.
The figure shows that when including spill-out the field profile becomes approximately three times broader. This effect becomes more pronounced for smaller $w$, and for $w=0.35$ nm the field is almost 100 times broader, which is due to the fact that the mode index is almost equal to the refractive index of bulk gold when including spill-out. 

The penetration length as a function of $w$ is shown in Fig. \ref{fig:H}(b) with and without spill-out at a wavelength of 775 nm. The blue line shows the case when neglecting spill-out, and for small $w$ the field is highly localized implying a very short penetration length, which was also observed in Fig. \ref{fig:H}(a). When including spill-out the penetration length shown by the red line for small $w$ becomes very high and diverges as shown in the inset. Thus, the field behaves almost like a plane wave in the limit $w\to 0$. On the other hand, when $w$ increases the penetration length becomes almost the same with and without spill-out. Then, at a first glance, the effect of spill-out seems to be negligible for gaps of a few nm, but as was found in the previous section significant absorption takes place in the gap region 0.15 nm from the interfaces. It is astonishing that even though the field penetration into the gold surfaces is almost the same with and without spill-out for few-nm gaps the absorption is still much higher when including spill-out. 
The horizontal green line in Fig. \ref{fig:H}(b) shows the penetration length of an SPP bound to a single interface between gold and air \cite{nanooptik}. When the gap becomes sufficiently wide the gap plasmon behaves almost as a single SPP bound to a gold surface, and the penetration lengths shown by the red and blue lines are found to converge to the horizontal green line for large $w$ (not shown).  

\begin{figure}[!h]
\begin{center}
\includegraphics[width=8cm]{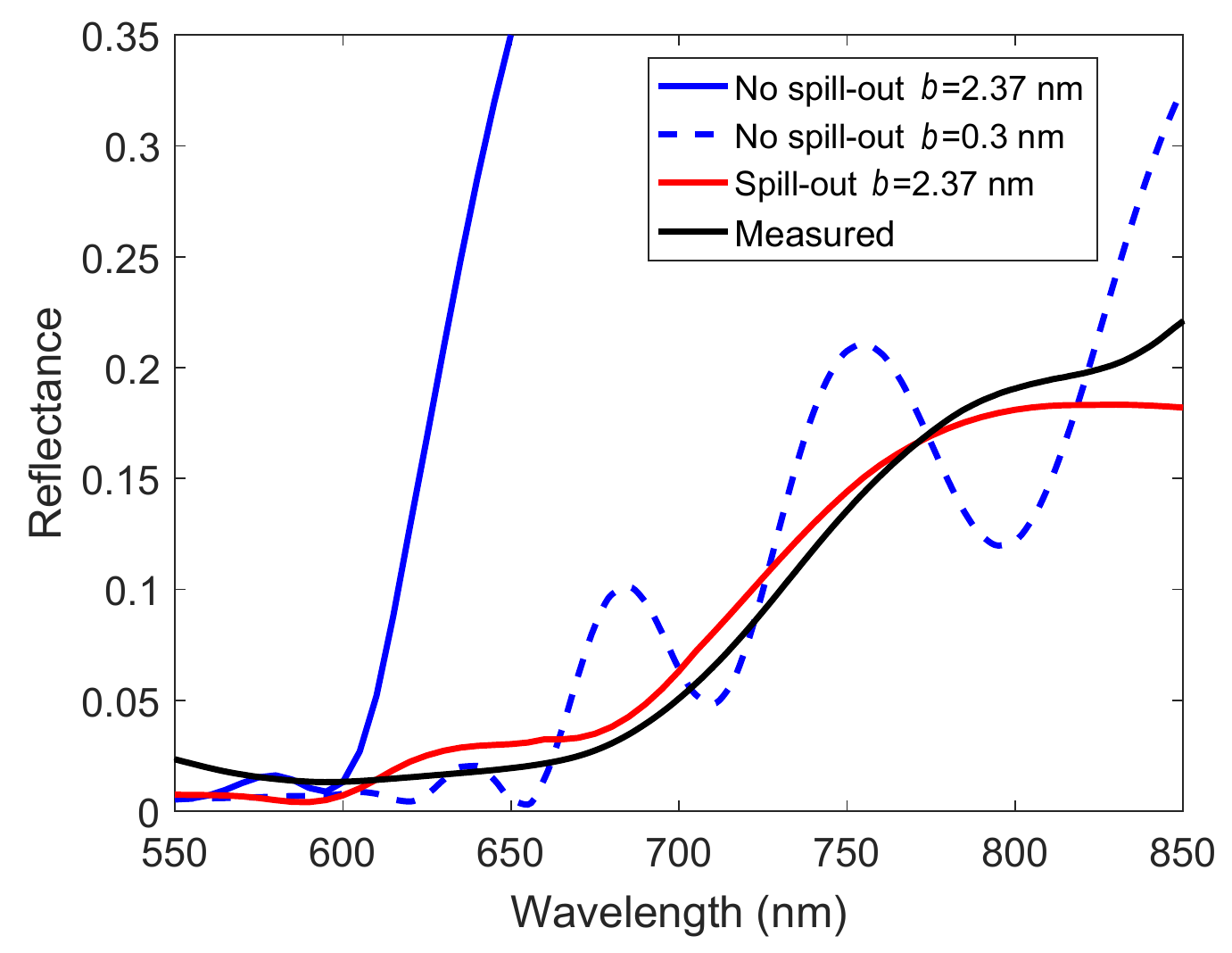}
\end{center}
\caption{Comparison between the calculated reflectance with and without spill-out and the measured reflectance from Ref. \citenum{black_gold} shown in black. Spill-out is included in the calculated reflectance shown by the red line and neglected in the calculated reflectance shown by the blue lines. The groove height is 450 nm, the top width is 240 nm, and $b$ denotes the bottom width. }
\label{fig:R_comparison}
\end{figure}

The bottom width $b$ of an ultrasharp groove is impossible to measure precisely. In the calculations in Refs. \citenum{black_gold,optics_multiple}, where spill-out was neglected, the bottom width was set to 0.3 nm, which is close to the gold atom diameter. This was necessary in order to obtain a reflectance comparable to measured values. However, it was found in Fig. \ref{fig:R} that the reflectance from an ultrasharp groove array with this bottom width is significantly lower when spill-out is included. It is therefore investigated in Fig.\ref{fig:R_comparison}, which impact the bottom width has on the calculated reflectance of a 450 nm deep ultrasharp groove array with top width 240 nm. For bottom widths below 1 nm the reflectance is always below 1.3\% (not shown), being thus still significantly lower than both measured and calculated values in Refs.\citenum{black_gold,optics_multiple}. By adjusting $b$ to minimize the root mean square error between calculated and measured reflectance, it is found that the bottom width $b=$ 2.37 nm gives the best reflectance. The calculated reflectance when including spill-out for this $b$ is shown by the red line Fig. \ref{fig:R_comparison}, and is in excellent agreement with the measured reflectance from Ref. \citenum{black_gold} shown in black in the same figure.

The blue dashed and solid lines in Fig. \ref{fig:R_comparison} show the calculated reflectance when neglecting spill-out for bottom widths of 0.3 nm and 2.37 nm, respectively. For $b$ = 2.37 nm the reflectance is now much higher than the measured reflectance. The result for 0.3 nm, also shown in Fig. \ref{fig:R}, is similar in magnitude to the measured reflectance shown in black but the oscillations in the calculated reflectance spectrum are clearly not present in the measured reflectance. On the other hand such oscillations are not present in the calculation that includes spill-out, in which case a much better agreement with the measured reflectance is obtained. 

In Ref. \citenum{black_gold}, the reflectance has been measured for several different fabricated arrays of ultrasharp grooves. In the theoretical calculations performed in this paper and in Refs. \citenum{black_gold,optics_multiple} it has been assumed that all the grooves in the periodic array are identical. This is extremely hard to guarantee in practice when fabricating the grooves, and SEM and optical microscope images of the arrays of grooves in Ref. \citenum{black_gold} also show that there are minor variations. The fact that the fabricated groove arrays are not perfectly periodic may explain some of the deviation between the calculated reflectance including spill-out and the measured reflectance. 

\section{Conclusion} \label{sec:conclusion}
Using a fully quantum mechanical approach, the properties of gap plasmons in ultranarrow metal gaps have been investigated. Electron spill-out is found to play a crucial role for both plasmon propagation and reflectance from ultrasharp groove arrays. In these geometries, a classical approach based on bulk optical properties leads to unphysically diverging mode indices in the limit of vanishing gap width. We demonstrate, however, that divergencies are avoided when spill-out is taken into account. Importantly, spill-out also has a great impact on gaps of a few nm, since power is strongly absorbed in the gap region 1-2 Å from the interfaces. As a consequence, calculated reflectance spectra are in excellent agreement with measured reflectance spectra for ultrasharp groove arrays.

\section*{Acknowledgement}
This work is supported by Villum Kann Rasmusen (VKR) center of excellence QUSCOPE.

\bibliography{mybib} 

\end{document}